\documentclass{ws-procs9x6}
\usepackage{bm}
\usepackage{hyperref}
\usepackage{amsmath,amssymb,amsfonts,slashed,graphicx,float}
\renewcommand{\vec}[1]{{\mathbf{#1}}}
\newcommand{\beq}{\begin{eqnarray}}
\newcommand{\eeq}{\end{eqnarray}}
\usepackage{amsmath}
\usepackage{amssymb}
\usepackage[paper=letterpaper,margin=1in]{geometry}
\usepackage{braket}
\usepackage{upgreek}

\renewcommand{\vec}[1]{\boldsymbol{#1}}

\def\a{\alpha}

\def\del{\partial}

\usepackage{tikz}

\begin{document}

\title{Beyond Particles: Unparticles in Strongly Correlated Electron Matter}
\author{ Philip W. Phillips }

\address{Department of Physics and Institute for Condensed Matter Theory,
University of Illinois
1110 W. Green Street, Urbana, IL 61801, U.S.A.}

\date{\today}

\begin{abstract}

I am concerned in these lectures with the breakdown of the particle concept in strongly correlated electron matter.  I first show that the standard procedure for counting particles, namely Luttinger's theorem, breaks down anytime pole-like excitations are replaced by ones that have a divergent self-energy.  Such a breakdown obtains in electronic systems whose pole-like excitations do not extend to the edge of the Brillouin zone, as in Fermi arcs in the cuprates.   Since any non-trivial infrared dynamics in strongly
correlated electron matter must be controlled by a critical fixed point, unparticles are the natural candidate to explain the presence of charged degrees of freedom that have no particle content.  The continuous mass formulation of
unparticles is recast as an action in anti de Sitter
space.  Such an action serves as the generating functional for the propagator. This mapping fixes the scaling dimension of
the unparticle to be $d_U=d/2+\sqrt{d^2+4}/2$ and ensures that the
corresponding propagator has zeros with $d$ the spacetime dimension of
the unparticle field.  The general {\it dynamical} mechanism by which bulk operators, such as the Pauli term, couple to the scaling dimension of the boundary operator and thereby lead to a vanishing of the spectral weight at zero energy is reviewed in the context of unparticles and zeros.  
The analogue of the BCS gap equations with unparticles indicates that the transition temperature increases as the
attractive interaction strength decreases, indicating that unparticles
are highly susceptible to a superconducting instability.  

\end{abstract}

\keywords{cuprates, unparticles, anti de Sitter}
\bodymatter

\section{Introduction}

In these lectures, I will focus on strong electron correlations as they play out in the normal state of the cuprates.   The ultimate goal is to understand the departures from the standard theory of metals in terms of a new fixed point, namely one that describes strong coupling physics.   While there are obvious features of the normal state that suggest that Fermi liquid theory breaks down, such as $T-$ linear resistivity, my starting point for approaching this problem will be the onset of Fermi arc formation\cite{norman98,dessau96,dessau,KingP11,YangH11} as the Mott insulator is doped.  My focus on Fermi arcs as a window into the origin of non-Fermi liquid behavior in the cuprates is quite simple.  As I will show, when arcs form,  the standard rule which relates the total charge count to the number of quasiparticles fails.  Consequently, Fermi arcs provide a glimpse into the breakdown of the particle concept in the normal state of the cuprates. I will then suggest that the unparticle construct of Georgi's\cite{georgi} provides a good description of the extra stuff that couples to the current but has no particle content.  I will then show that such scale-invariant matter has a natural mapping on to an action in anti de Sitter space and has a superconducting instability quite unlike the standard BCS picture. 

\section{Fermi arcs and the Failure of the Luttinger Count}

\begin{figure}
\centering
\includegraphics[width=6.5cm]{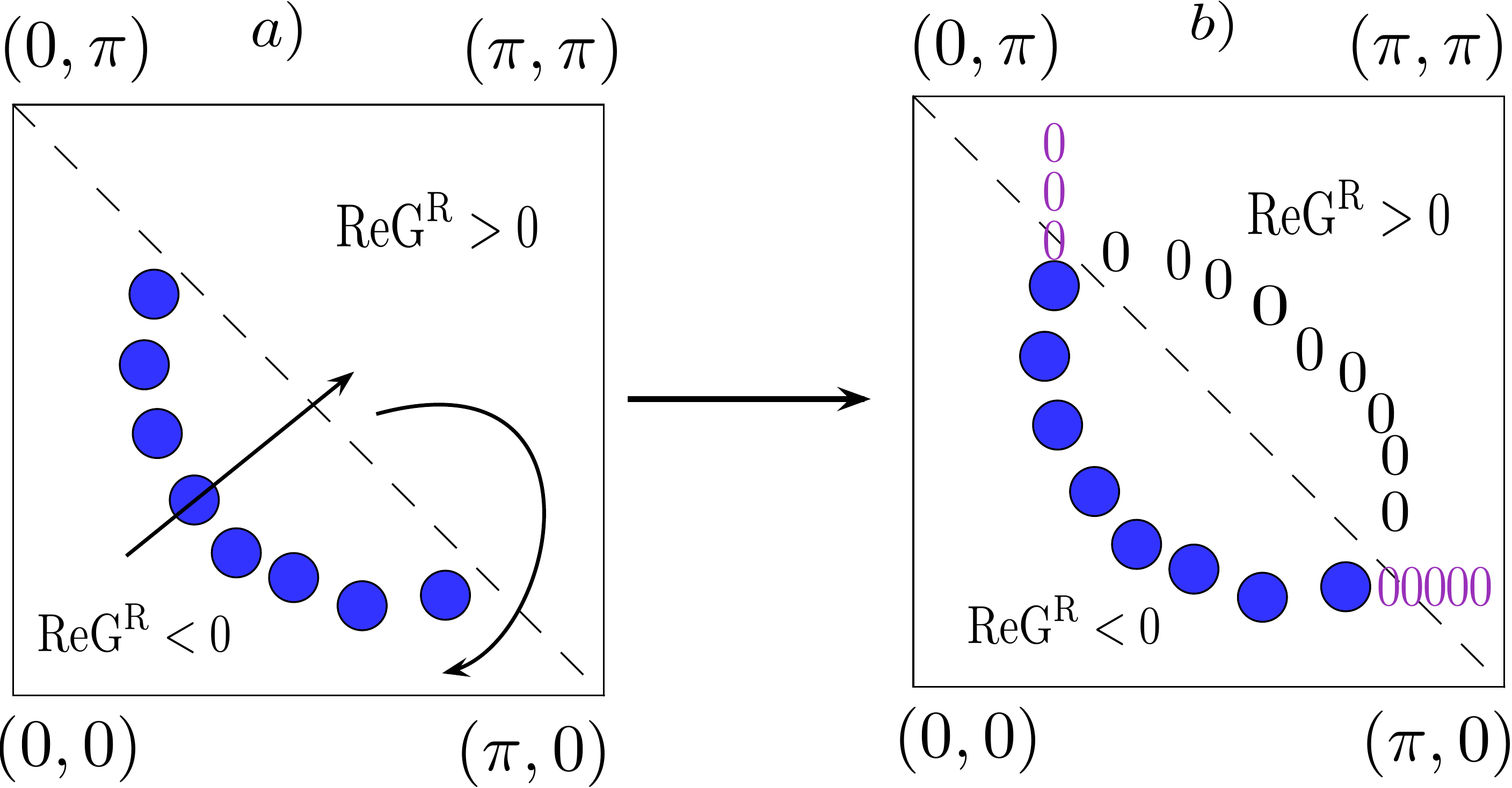}
\caption{(Color online) a.) Sign change of the real part of the retarded
  Green function in the presence of an arc of quasiparticle excitations, blue dots.   The path shown, through the Fermi arc and circumventing it on the return is problematic because the return path has no mechanism for picking up the sign change of the Green function.  b) Solution for maintaining the correct sign of the Green function.  A line of zeros must be present in one of two places:  1) either on the backside of the arc or 2) somewhere between the end of the arc and the edge of the Brillouin zone. }
\label{zeropoint}
\end{figure}

As illustrated in Fig. (\ref{zeropoint}a), lightly doping a Mott insulator\cite{norman98,dessau96,dessau,KingP11,YangH11} results in a locus of quasiparticle excitations  in momentum space which does not form a closed surface.  The quasiparticles, represented by solid dots in Fig. (\ref{zeropoint}a), form an arc in momentum space which does not extend to the Brillouin zone edge.   Because the quasiparticle excitations correspond to poles in a single-particle Green function, 
\beq
G^R=\frac{Z_p}{\omega-\varepsilon_p-\Sigma(\omega, \vec p)},
\label{gf}
\eeq
there are two distinct ways of viewing the termination of the arc.  Either poles still exist where the would-be arc terminates but the spectral weight, $Z_p$ is too small to be detected experimentally or there is no pole at all.  The first is rather traditional and requires no modification of the standard theory of metals.  That is, the former is simply a limitation on experiment.  However, such is not the case for the latter.  To illustrate the difficulty, consider the path shown in Fig. (\ref{zeropoint}a).  On either side of the Fermi arc or the extreme case of a nodal metal in which case there is only a quasiparticle at $(\pi/2,\pi/2)$\cite{dessau}, the real part of the Green function must change sign.  However, the path depicted in Fig. (\ref{zeropoint}a) passes through the Fermi arc but returns without encountering a pole and hence has no mechanism for incurring the expected sign change.  The only possible way to end up with the correct sign of the real part of the Green function is through a line of zeros of the single-particle Green function as depicted in Fig. (\ref{zeropoint}b).  The zero surface can either be on the back-side of the arc or somewhere between the end of the arc and the edge of the Brillouin zone.   Consequently, numerous proposals\cite{yrz,kotstan,konik2006,zeros2,NormanM07} for the pseudo gap state of the cuprates have stressed that zeros of the retarded Green function (more rigorously, ${\rm Det}\Re G^R(E=0,\vec p=0$)) must be included if Fermi arcs are to be modeled.  

That zeros of the single-particle Green function represent a breakdown of the particle concept can be seen from Eq. (\ref{gf}).  The only way for the Green function to vanish in the absence of a pole is if the denominator diverges.  This requires a divergent self energy, $\Sigma$.   Such a divergence tells us that the starting point of particles with well-defined energy, momentum, or mass is no longer valid.  Nonetheless, many have argued\cite{dzy,yrz,zeros2,konik2006} that they must be included in the count,
\beq
\label{nlutt}
n=2\sum_{\vec p}\theta(\Re G^R(\omega=0,\vec p)),
\eeq
of the particle density. This seemingly innocuous expression is a consequence of what is known as Luttinger's theorem\cite{luttinger}, a prescription for counting particles.
 Because $\theta$ is the heavy-side step function, this equation has contributions anytime the real part of the Green function is positive.  This occurs via a pole or a zero crossing.  Poles  count quasiparticles and hence it is natural that they should be included in the particle count.  Should only poles be present, as in a Fermi liquid, then the physical content of Eq. (\ref{nlutt}) is self evident.  However, this equation also includes zero crossings where the particle concept breaks down.  Nonetheless, they are present in the Luttinger count of the particle density as pointed out explicitly first in the classic book by Abrikosov, Gorkov, and Dzyaloshinski\cite{agd} ( Eq. (19.16) and the paragraph below it) and others \cite{dzy,konik2006,zeros2}.  
 
 There is a simpler way of counting charges.  Simply integrate the density of states
 \beq
 n=2\int_{-\infty}^{\mu}N(\omega)
 \eeq
 over all the occupied states.  As a sum over all energy scales, this formula for the charge count has both high-energy ($\rm UV$) and low-energy ($\rm IR$) physics.  Hence, what Eq. (\ref{nlutt}) accomplishes is a reduction of the charge density to a zero-energy ($\omega=0$) surface.  That is, the charge density has been reduced to $\rm IR$ physics alone and as a consequence, all the charge content has a particle interpretation.
 
 If Eq. (\ref{nlutt}) is correct, then it is truly extraordinary because it would apply in cases where the self energy vanishes and cases in which it diverges.  A classic example of a divergent self-energy is the Mott insulator\footnote{Dzyaloshinkskii\cite{dzy} has maintained (see discussion after Eq. (9)) that in a Mott insulator, the self energy does not diverge but nonetheless there are still zeros, in the sense that $\rm Det Re G^R(\omega=0,\vec p)=0$ in a Mott insulator.  These statements are contradictory.}.  Consider the Kramers-Kronig
 \beq
 \label{GRint}
 \Re G^R(\omega=0,\vec p)=\frac{1}{\pi}\int_{-\infty}^{\infty} \frac{\Im G^R(\omega',\vec p)}{\omega'} d\omega'
 \eeq
 relationship for the real part of the retarded Green function.   A Mott insulator is a gapped system in which the parts of the spectral function below and above the chemical potential have the same band index.  As a result, there is no change in the band index upon integration over all frequency.  Since $\rm Im G^R$ is of a fixed sign,  the integrand in Eq. (\ref{GRint}) can change sign in the range of integration.  Consequently, the integral can vanish.  A zero of the single-particle Green function then is a balancing act of spectral weight above and below the chemical potential weighted with the factor of $1/\omega$.   As a result, the zero surface is strongly dependent on where the chemical potential is located.  While this might seem to make sense in the context of the particle density being determined by the location of the chemical potential, there is a subtlety here in that even for an incompressible system, in which case the chemical potential can be placed anywhere in the gap, the location of the zero surface can change, thereby changing the particle densty\cite{rosch}.  Physically, however, nothing changes when the chemical potential is moved in the gap.  This problem was first pointed out by Rosch\cite{rosch} and can be illustrated with the atomic limit of the Hubbard model $SU(2)$ Hubbard model in which the Hamiltonian is simply $U n_{\uparrow}n_{\downarrow}$.  The retarded Green function for this problem
 \beq
 G^R(\omega)=\frac{1}{\omega+\mu+\frac{U}{2}}+\frac{1}{\omega+\mu-\frac{U}{2}}
 \eeq
 is a sum of the two poles that constitute the lower and upper Hubbard bands.  It is instructive to write both terms over a common denominator,
 \beq
 G^R(\omega)=\frac{1}{\omega+\mu+U/2-\Sigma_{\rm loc}(\omega)},
 \eeq
 which allows us to identify the self-energy,
 \beq
 \Sigma(\omega)=\frac{U}{2}+\left(\frac{U}{2}\right)^2\frac{1}{\omega+\mu}.
 \eeq
 Clearly when $\omega=-\mu$, the self energy diverges, thereby leading to a zero of the Green function in contrast to the statement by Dzyaloshinskii\cite{dzy} that the self-energy does not diverge in a Mott insulator.   Evaluating at $\omega=0$ and substituting the result into the Luttinger theorem leads to
 \beq
 n=2\Theta\left({\frac{\mu}{\mu^2-\left(\frac{U}{2}\right)^2}}\right)
 \eeq
 the expression that should yield the particle density.  The Mott insulator corresponds to $n=1$ and this formula should yield this result for any value of the chemical potential satisfying $-U/2<\mu<U/2$.  However, it is evident that there is a problem here:  1) for $-U/2<\mu<0$, n=2, 2) $\mu=0$, $n=1$, and finally 3) for $0<\mu<U/2$, $n=0$.  Hence, unlike poles, zeros are not a conserved quantity.  This does not mean they are devoid of physical content.  Zeros are the key indicator that high and low energies are mixed as in the Mott problem laid plain by Eq. (\ref{GRint}).  So what to make of this result?  It might be argued\cite{farid,rosch2} that one could take care of the degree of freedom with the chemical potential by using the limiting procedure, $\lim_{T\rightarrow 0}\mu(T)$ which would uniquely fix the chemical potential at $T=0$.  For the $\rm SU(2)$ atomic limit of the Hubbard model, this limiting procedure places the chemical potential at the particle-hole symmetric point, $\mu=0$, in which case $n=1$.  Does this fix all the problems?
 
While the importance of this limiting procedure can certainly be debated\cite{farid,rosch2}, it would be advantageous to consider a model which even with this limiting procedure, the Luttinger count is violated.  The generalization to a non-degenerate ground state will be considered a little later.  The model we consider is a generalization of the atomic limit of the Hubbard model,
 the $\rm SU(N)$ 
 \beq\label{nluttb}
 H=\frac{U}{2}(n_1+n_2+\cdots n_N)^2,
 \eeq
 to $N$ degenerate flavors of fermions and hence has $SU(N)$ symmetry.   As will be seen, it is imperative that this symmetry be in tact because the basic physics of Mott insulation fails when the spin rotation symmetry is broken.  Exact computation of the Green function\cite{kiaran} reveals that the equivalent statement of Luttinger's theorem for the $SU(N)$ model becomes
 \beq\label{nluttb}
 n=N\Theta(2n-N).
 \eeq
 Consider the case of $n=2$ and $N=3$.  This equation implies that $2=3$ and hence is false.  In fact, any partially filled band with $N$ odd leads to a violation of Eq. (\ref{nlutt}).  That
Eq. (\ref{nluttb}) actually reduces to the correct result for the $\mathrm{SU}(2)$ case is entirely an accident because the $\Theta$ function only takes on values 
of $0$, $1/2$, or $1$.  This problem still persists even if the degeneracy of the state in the atomic limit is lifted by a small hopping $t=0^+$.  Note this perturbation preserves the $SU(N)$ symmetry and hence is fundamentally different from lifting the degeneracy via a magnetic field which would break the $SU(N)$ symmetry as proposed recently as a rebuttal to our work\cite{farid2}.  In the limit $t\rightarrow 0$, $\mu=0$, and  $T=0$, the Green function becomes
 \beq
G_{ab}(\omega)={\rm Tr}\left(\left[c_a^\dagger\frac{1}{\omega-H}c_b+c_b\frac{1}{\omega-H}c_a^\dagger\right]\rho(0^+)\right)\nonumber
\eeq
with $\rm Tr$ the trace over
 the Hilbert space and $\rho(0^+)$ the density matrix.  Here $c_a$ is the creation operator for a fermion of flavor, $a$.  Our use of $\rho(0^+)$ is crucial here because $\rho(t=0^+)$
 describes a pure state whereas for $\rho(t=0)=\sum_u
 P_u|u\rangle\langle u|$ with probabilities satisfying $\sum_u
 P_u=1$ is a mixture of many
 degenerate ground states $|u\rangle$.  Our use of $\rho(0^+)$ allows
 us to do
 perturbation theory in $t$.  For $t\rightarrow 0$, the intermediate
 states have energy $U$ or $0$ and hence we can safely pull $\omega-H$
 outside the trace.  Noting that $\{c_a^\dagger,c_b\}=\delta_{ab}$, we obtain that
\beq\label{defcon5}
G_{ab}(\omega)=\frac{\omega\delta_{ab}-U\rho_{ab}}{\omega(\omega-U)},
\eeq
where we have introduced $\rho_{ab}={\rm Tr}\left(  c_a^\dagger c_b
  \rho(0^+)\right)   =  \langle u_0 |  c_a^\dagger c_b | u_0 \rangle$,
$|u_0\rangle$ the unique ground state.  Now comes the crucial point. Consider $N=3$. If the degeneracy between the iso-spin states were lifted in the starting Hamiltonian, as in a recent model\cite{farid2} put forth as a possible rebuttal to our claim that Eq. (\ref{nluttb}) represents a counterexample to the Luttinger claim, only a single one of the iso-spin states would be possible
and $\rho_{ab}(0^+)={\rm diag}(1,0,0)\ne \rho_{ab}(t=0)$.  However, turning on a
hopping matrix element places no restriction on the permissible
iso-spin states.  Consider a rotationally-invariant spin singlet state
on a three-site system, the unique ground state.  
We have that $\rho_{ab}(0^+)=1/3
{\rm diag}(1,1,1)=\rho_{ab}(t=0)$. For this unique
ground state, Eq. (\ref{defcon5}) has a zero at $\omega=U/3$ whereas
 $\lim_{T\rightarrow 0}\mu(T)=U/2$.  Consequently, the Luttinger
count fails to reproduce the particle density. Specifically,
Eq. (\ref{defcon5}) implies that $1=0$!  The key point is that as long as the perturbation
which lifts the degeneracy does not break $SU(N)$ symmetry, then
$\rho_{ab}(t=0^+)=\rho_{ab}(t=0)$ and Eq. (\ref{defcon5})
survives and the recent criticism\cite{farid2} does not. 
 
 The crux of the problem is that the
Luttinger-Ward (LW) functional strictly does not exist when zeros of
the Green function are present.  Consider the LW functional, defined by
\begin{subequations}
\begin{align}
\delta I[G]&=\int d\omega \Sigma\delta G\\
I[G=G_0]&=0
\end{align}
\end{subequations}
which was used by Luttinger\cite{luttinger} to show that the integrand
of $I_2$ is a total derivative. Because $\Sigma$ diverges for some
$\omega$ when $G$ is the total Green function, it is not possible to
integrate the defining differential expression in the neighborhood of
the true Green function, and therefore the LW functional does not
exist.  Consequently, there is no Luttinger theorem and
Eq. (\ref{nlutt}) does not represent the density of a fermionic system
because zeros of the Green function must be strictly excluded, a
model-independent conclusion.  In fact, Dzyaloshinskii\cite{dzy} has already pointed out that when $\Sigma$ diverges, the Luttinger count fails.  However, he mistakenly assumes  that such a divergence is not intrinsic to a zero of the Green function and hence is irrelevant to Mott physics.  As we have seen here, this is not the case.
 
\begin{figure}[h!]
\begin{center}
\includegraphics[width=3.0in]{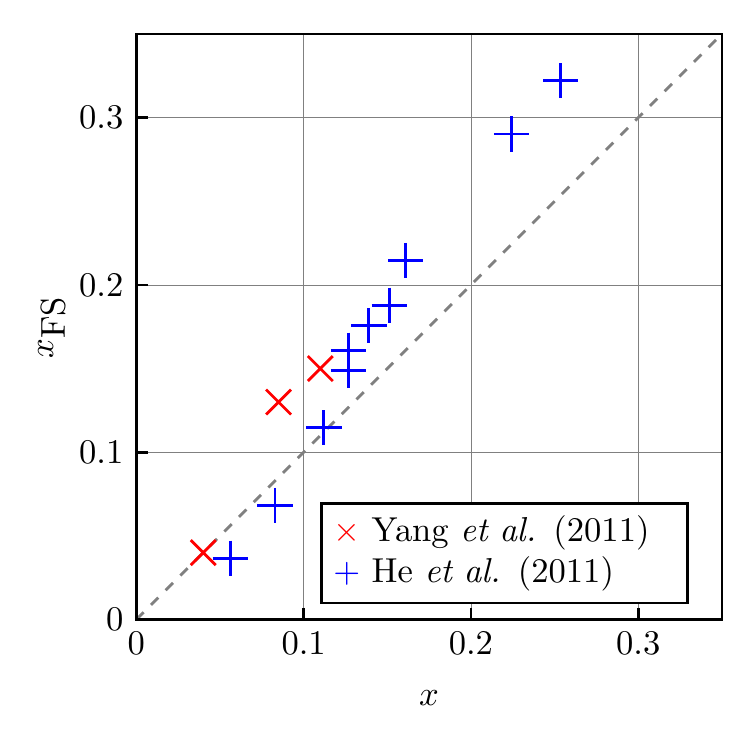}
\caption{Apparent doping $x_{\mbox{FS}}$ inferred from the Fermi surface reconstruction as a function of the nominal doping $x$ in LSCO and Bi-2212. }
\label{fig:ARPESplot}
\end{center}
\end{figure}

Experimentally, Fermi arc formation leads to a violation of Luttinger's theorem.  Shown in
Fig. (\ref{fig:ARPESplot}) is a plot of the area enclosed by the locus of k-points for which there is a maximum
in the spectral function in La$_{2-x}$Sr$_x$CuO$_2$\cite{shen} ($+$
plot symbol) and Bi$_2$Sr$_2$CaCu$_2$O$_{8+\delta}$\cite{YangH11}
($\times$ plotting symbol) as a function of the nominal doping level
in the pseudogap regime. Although the maxima in the spectral
function form an arc as there are zeros present on the opposite side,
$x_{\rm FS}$ was extracted by simply closing the arc
according to a recent proposal\cite{yrz} for
Bi$_2$Sr$_2$CaCu$_2$O$_{8+\delta}$\cite{YangH11} (Bi-2212) and for
La$_{2-x}$Sr$_x$CuO$_2$\cite{shen} (LSCO) by
determining the large Fermi surface ($1-x$) defined by the $k_F$ measured
directly from the momentum-distribution curves and then
subtracting unity.  Hence, the key
assumption that is being tested here in this definition of $x_{\rm
  FS}$ is that each doped hole corresponds to a single $k-$state.  A typical uncertainty in these experiments
is $\pm 0.02$.  Even when this uncertainty is considered, the
deviation from the dashed line persists indicating that one hole does
not equal one k-state and hence a fundamental breakdown of the elemental
particle picture in the cuprates.

 \section{Unparticles and anti de Sitter Spacetime}
 
\begin{figure}
\begin{center}
\includegraphics[width=3.0in]{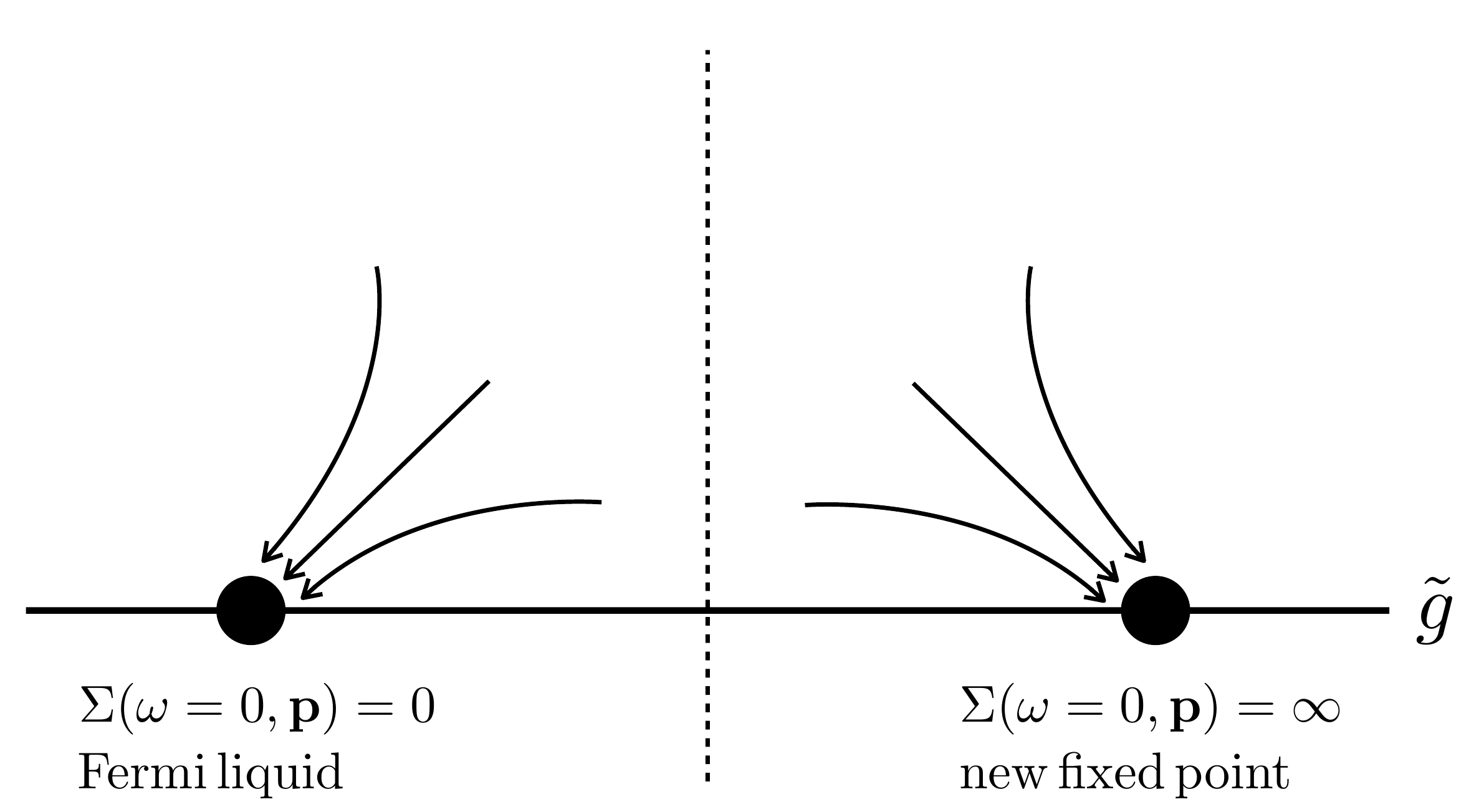}
\caption{  Heuristic renormalization group flow for the Fermi liquid fixed point in which the self-energy is zero or negligible in the IR and
one in which zeros of the single-particle propagator appear.  Since the self energy diverges in the latter, the excitations which appear here are not adiabatically connected to the Fermi liquid fixed point.  The breakdown of the particle concept at the new fixed point suggests an unparticle picture is valid.  The horizontal axis represents the strength of the coupling constant.}
\label{rgpict}
\end{center}
\end{figure}

What I have shown so far is that there is some charged stuff which has no particle interpretation when Luttinger's theorem breaks down.  So what is the extra stuff?  The dichotomy in Fig. (\ref{rgpict}) represents the current conundrum.
Any
excitations that arise from such a divergence of the self energy are clearly not 
adiabatically connected to the ones at the non-interacting or Fermi-liquid fixed point.
Consequently, if any new excitations emerge, they must arise
fundamentally from a new fixed point as illustrated in Fig. \ref{rgpict}. All fixed points have scale invariance.  
To illustrate scale invariance, consider the Lagrangian
\beq\label{lagrangian}
L=\frac12(\partial_\mu\varphi)^2+m^2\phi^2
\eeq
with a mass term.   Without the mass term, the scale transformation $x\rightarrow x/\Lambda$ produces a Lagrangian of the same form just multiplied by the scale factor $\Lambda^2$.  Hence, the Larangian without the mass term is scale invariant.  However, with the mass term,
scale invariance is lost.   For a theory with a coupling constant, that is a term of the form $g\varphi^4$, scale invariance at a critical point 
\beq\label{beta}
\beta(g)\equiv \frac{dg}{d\ln E}=f(g)=0,
\eeq
reflects an invariance of the coupling constant under a change in the energy scale. That the $\beta-$function is local in energy implies that renormalization group (RG) procedure should be geometrizable.  It is this insight that underlies the gauge-gravity duality\cite{maldacena} as we will see later in this section.

While it is difficult to establish the existence of a non-trivial
(non-Gaussian in the UV variables) IR fixed point of the Hubbard model (or any strongly
coupled model for that matter), the cuprates, Mott systems in general,  display quantum critical scaling\cite{kanoda,mc1,marel,valla,mfl}.
This suggests that it is not unreasonable to assume that a strongly coupled fixed point governs the divergence of the self energy in the Fermi arc problem.
Without knowing the details of the fixed point, what permits immediate quantitative progress here is that all critical
fixed points exhibit scale invariance and this principle
anchors fundamentally the kind of single-particle propagators that arise as pointed
out by Georgi\cite{georgi}.  If $\vec k$ is the 4-momentum, the candidate propagator must have an algebraic form, $(\vec k^2)^\gamma$.  However, such a  propagator cannot describe particles because particles are not scale invariant.  Scale invariant stuff has no particular mass, hence the name unparticles.  The most general form of the single-particle propagator,
\beq\label{gfunparticles}
G_U(\vec{k})&=& \frac{A_{d_U}}{2\sin(d_U\pi)}\frac{i}{(
  \vec{k}^2-i\epsilon)^{d/2-d_U}},\nonumber\\
 A_{d_U}&=&\frac{16\pi^{5/2}}{(2\pi)^{2d_U}}\frac{\Gamma(d_U+1/2)}{\Gamma(d_U-1)\Gamma(2d_U)},
\eeq
involves the scaling dimension of the unparticle field, $d_U$, as proposed by Georgi\cite{georgi}.   Here $ \vec k$ is the $d$-momentum where $d$ is the dimension of the spacetime in which the unparticle lives. Here, we adopt the notation where the diagonal entries of the metric are mostly positive. 

Within the unparticle
proposal, the spectral function should have the form
\beq\label{specfun}
A(\Lambda\omega,\Lambda^{\a_{ \vec{k}}} \vec{k})&=&\Lambda^{\alpha_A}A(\omega,\vec k),\nonumber\\
A(\omega,\vec{k}) &=& \omega^{\alpha_A}f_A\left(\frac{\vec k}{\omega^{\a_{\vec{k}}}}\right).
\eeq
We take $\a_A=2d_U-d$.
The scaling for a Fermi liquid corresponds to
$d_U=(d-1)/2$. Because of the constraints on unparticles, $d_U$ always
exceeds $d/2-1$ for scalar unparticles and $(d-1)/2$ for unfermions, where the rule to
obtain unfermions is to set $d \to d+1$.  We term a correlated system with
such scaling an un-Fermi liquid as the basic excitations are
unparticles. Un-Fermi liquids are non-Fermi liquids composed of
unparticles, whose propagator is given by the unfermionic analogue of
Eq. (\ref{gfunparticles}),
\beq\label{unf}
S_{U}\left(\vec k\right) & \sim &
\left(\vec k^{2}-i\epsilon\right)^{d_{U}-(d+1)/2}\times\nonumber\\
&&\left(\vec {k}\!\!\!/+\cot\left(d_{U}\pi\right)\sqrt{\vec{k}^{2}-i\epsilon}\right),\nonumber\\
\eeq
which contains a non-local mass term.
Un-Fermi liquids should not be construed as Fermi
liquids with poles at the unparticle energies.  Because the unparticle
fields cannot be written in terms of canonical ones, there is no sense
in which a Gaussian theory can be written down from which pole-like
excitations can be deduced.  Note that the scaling form, if it were
to satisfy any kind of sum rule, can only be a valid approximation
over a finite energy range.  

From where do unparticles come?  This can be illustrated with the continuous mass formalism\cite{cm1,deshe}.  The key idea here is that although unparticles have no particular mass, it is still possible to use the massive Lagrangian in Eq. (\ref{lagrangian}) to construct the propagator for unparticles.  The trick is to view unparticles as a composite with all possible mass\cite{cm1,deshe}.  We consider the action,
\beq\label{gunact}
S_{\phi}=\frac12\int_0^\infty f(m^2) dm^2\left[\int d^{4}\vec{x} \, \left(\partial_{\mu} \phi \partial^{\mu} \phi + m^2
  \phi^2\right)\right]\nonumber\\
\eeq
 in which the mass is explicitly integrated over with a distribution function of the form, $f(m^2)\propto (m^2)^{d_U-d/2})$ and hence $\varphi\equiv\varphi(x,m^2)$.  Because of the mass integration, any scale change in the Minkowski variables, $x^\mu$, can be absorbed into a redefinition of the mass integral and hence our action is scale invariant.  The propagator for this action,
 \beq
 \int^\infty_0 dm^2 \frac{(m^2)^{d_U-d/2}}{\vec p^2+m^2}\propto (\vec p^2)^{d_U-d/2},
 \eeq
has the algebraic form describing unparticles.  An identical analysis applies to Dirac fields as well\cite{deshe}. Hence, the continuous-mass formalism lays plain that unparticles should be thought of as scalar or Dirac fields with no particular mass.  

While it is common\footnote{For example, this has been used explicitly in the derivation of the unparticle propagator by
Deshpande and He\cite{deshe}.} to relate
the emergent unparticle field directly to the scalar field $\phi$
using a different mass-distribution function $g(m^2)$ through a relationship of the form
\beq\label{uphiphi}
\phi_U(\vec{x})=\int_0^\infty dm^2 \phi(\vec{x},m^2)B(m^2),
\eeq
this is not correct
as it would imply that the unparticle field $\phi_U$ has a particle
interpretation in terms of the scalar field $\phi$.  For
example, $\phi_U$ would then obey a canonical commutator and the resultant unparticle propagator could be interpreted as that of a Gaussian theory.
We demonstrate this explicitly in the Appendix.  In actuality, the unparticle field should not be a sum of $\phi(x,m^2)$s, which are independent functions of
mass, but rather should involve some unknown product of the particle fields\cite{georgi}.   Consequently, unparticle physics cannot be accounted for by a Gaussian theory.  In fact, as pointed out by Georgi\cite{georgi}, unparticles should be thought of as a composite consisting of $d_U/2$ massless particles.
 
\begin{figure}
\begin{center}
\includegraphics[width=3.0in]{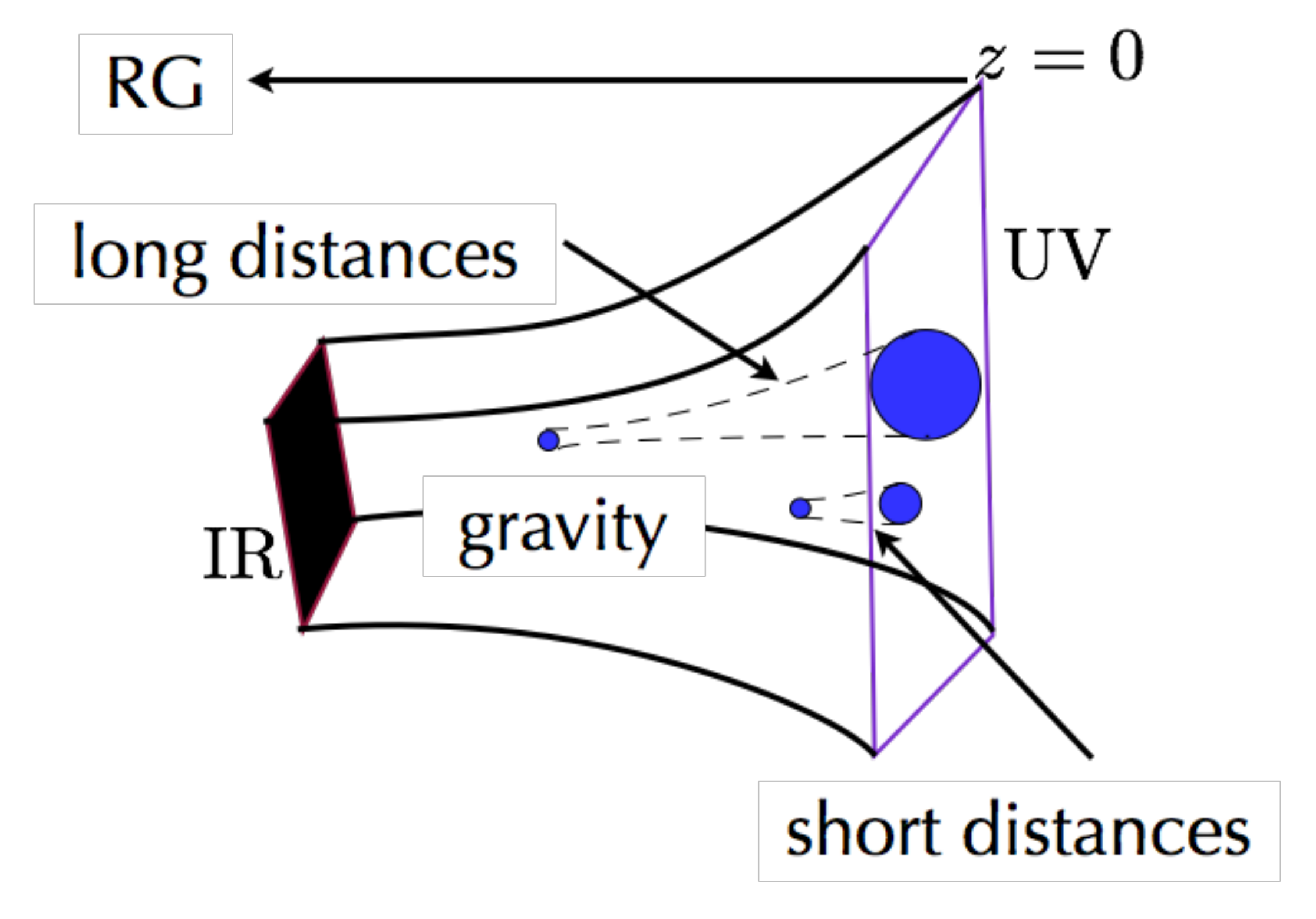}
\caption{  Geometrical representation of the key claim of the gauge-gravity duality.  A strongly coupled field theory lives at the boundary, at the UV scale.  The horizontal r-direction (the extra dimension in the gauge-gravity duality) represents the running of the renormalization scale.  This is illustrated by the two projections at different values of $z$ in the space-time.  Because the spacetime is asymptotically hyperbolic, larger values of $r$ lead to a larger projection of the boundary theory and hence full running of the renormalization scale amounts to the construction of the infrared (IR) limit of the original strongly coupled UV theory.  Charging the bulk gravity theory amounts to placing a black hole at $z=\infty$.}
\label{rgegr}
\end{center}
\end{figure}

However, it is the integration over mass that gives unparticles a natural interpretation within the gauge-gravity duality\cite{maldacena}.  The key claim of this duality is that some strongly coupled conformally invariant field theories in d-dimensions are dual to a theory of gravity in a $d+1$ spacetime that is asymptotically anti de Sitter
 \beq\label{adsmetric}
 ds^2=\frac{R^2}{z^2}\left(\eta_{\mu\nu}dx^\mu dx^\nu+dz^2\right).
 \eeq
This spacetime is invariant under the transformation $x_\mu\rightarrow \Lambda x_\mu$ and $z\rightarrow z\Lambda$ and hence satisfies the requisite symmetry (not the full symmetry of the conformal group) for the implementation of the gauge-gravity duality.  The requirement on the AdS radius is that it exceed the Planck length, $\Lambda_P$.  Because the AdS radius and the coupling constant of the boundary theory are proportional, the requirement, $R\gg \Lambda_P$,  translates into a boundary theory that is strongly coupled\cite{maldacena}.  Our current understanding of the extra dimension on the gravity side is that it represents the renormalization direction, that is the flow in the energy scale.   The scale change, $x_\mu\rightarrow \Lambda x_\mu$ increases the radial coordinate, $z\rightarrow z\Lambda$.   Consequently, moving into the bulk of the geometry increases the corresponding projection onto the boundary, as depicted in Fig. (\ref{rgegr}).  Hence, the limit of $z=\infty$ represents the full low-energy or IR limit of the strongly coupled theory, thereby providing a complete geometrization of the RG procedure.  That is $\rm RG=GR$.  Ultimately it is the locality of the $\beta-$ function with respect to energy that  is responsible for this geometrical interpretation of RG.   

Establishing a hard connection rather than a purely heuristic one\cite{Stephanov:2007uq,Cacciapaglia:2008kx} between unparticles and the gauge-gravity duality is straightforward within the continuous mass formalism.  The continuous mass formalism involves an integration over mass.  However, mass is energy and hence can be replaced by an inverse length.  Since there is already an extra dimension floating around in the continuous mass formalism, we can formalize\cite{unparticles} this by making the replacement $m\rightarrow 1/z$ in Eq. (\ref{gunact}).  Letting $f(m^2)=a_\delta (m^2)^\delta$ and remembering that $dm^2\rightarrow -1/z^3$ introduces an extra factor of $z^{5+2\delta}$ into the
continuous mass Lagrangian as can be seen from
\beq
\mathcal{L}&=&a_\delta\int_0^\infty
dz\frac{2R^2}{z^{5+2\delta}}\left[\frac{1}{2}\frac{z^2}{R^2}\eta^{\mu\nu}(\del_\mu
  \phi)( \del_\nu \phi)+\frac{\phi^2}{2R^2}\right]. \nonumber\\
\eeq
All the factors of $z$ and the radius, $R$, can be accounted for by using the AdS metric, Eq. (\ref{adsmetric}).
 Consequently, we have proposed\cite{unparticles} that the correct starting point for the unparticle construction
is  the action on AdS$_{5+2\delta}$
\beq
S = \frac{1}{2} \int d^{4+2\delta}\vec{x} \, dz \, \sqrt{-g} \left(\partial_a \Phi \partial^a \Phi + \frac{\Phi^2}{R^2} \right).
\eeq
where 
 $\sqrt{-g}=(R/z)^{5+2\delta}$, and
$\phi\rightarrow (a_\delta 2)^{-1/2}R^{3/2}\Phi$.  All of the factors of  $z^{5+2\delta}$ and the $z^2$ in the gradient
terms appear naturally with this metric.  According to the gauge-gravity duality, the on-shell action then becomes the
generating functional for
unparticle stuff which lives in an effective dimension of $d= 4+2\delta$. Therefore, we would like to have $\delta \leq 0$.  We have of course introduced extra dynamics in the $z-$ direction.  This will be eliminated by choosing the solutions which are non-normalizable at the boundary.  This will fix the scaling dimension of the unparticle field.  

We have shown that the 
unparticle propagator\cite{unparticles} falls out of this construction.  To see this, we start from the equation of motion,
\beq
z^{d+1} \, \partial_z \left(\frac{\partial_z \Phi}{z^{d-1}}\right) + z^2 \, \partial_{\mu} \partial^{\mu} \Phi - \Phi = 0.
\eeq
For spacelike momenta $k^2 > 0$, the solutions are identical to those of Euclidean AdS. The solution that is smooth in the interior is given by
\beq
\Phi(z,\vec{x}) = \int \frac{d^{d}\vec{k}}{(2\pi)^{d}} e^{i \vec{k}\cdot \vec{x}} \frac{z^{\tfrac{d}{2}} K_{\nu} (kz)}{\epsilon^{\tfrac{d}{2}} K_{\nu} (k\epsilon)} \, \tilde{\Phi}(\vec{k}),
\eeq
where $\vec{k}$ is a $d$-momentum transverse to the radial $z$-direction and 
\beq
\nu = \frac{\sqrt{d^2 + 4}}{2}. 
\eeq
We note that this solution decays exponentially in the interior and thus, even though it is a $z$-dependent solution, one can think of $\Phi$ as localized at the boundary $z=\epsilon \rightarrow 0$.  Hence, we have eliminated the unwanted dynamics in the bulk. Here, we have explicitly cut off the AdS geometry to regularize the on-shell action
\beq
S &=& \frac{1}{2} \int d^{d}\vec{x} \, g^{zz} \sqrt{-g} \,\Phi(z,\vec{x}) \partial_z \Phi(z,\vec{x}) \Big|_{z= \epsilon} \nonumber \\
&=& \frac{1}{2} \frac{R^{d-1}}{\epsilon^{d-1}} \int
\frac{d^{d}\vec{p}}{(2\pi)^{d}}\frac{d^{d}\vec{q}}{(2\pi)^{d}}
\, (2\pi)^{d} \delta^{(d)}(\vec{p}+\vec{q})\nonumber\\
&& \tilde{\Phi}(\vec{p}) \frac{d}{d\epsilon} \left(\log \left[\epsilon^{\tfrac{d}{2}} K_{\nu}(p\epsilon)\right]\right) \tilde{\Phi}(\vec{q}). \nonumber \\
\eeq
Interpreting this as a generating functional for the unparticle field
$\Phi_U$ living in a $d$-dimensional spacetime, we
can then read the (regulated) 2-point function, which scales as
$p^{2\nu}$. We can then  analytically continue to the case of timelike momenta, which corresponds
 to choosing the non-normalizable solution to the bulk equation of motion. This analytically continued solution will also be localized at the boundary. 

The correlation function for two unparticle fields in real space is then given by
\beq
\langle \Phi_U (\vec{x}) \Phi_U(\vec{x}') \rangle = \frac{1}{|\vec{x} -\vec{x}'|^{2d_U}},
\eeq
where 
\beq\label{du}
d_U = \frac{d}{2} + \frac{\sqrt{d^2 + 4}}{2} > \frac{d}{2},
\eeq
and $d$ is the dimension of the spacetime the unparticle lives
in.  We note that in this construction, there is only one possible scaling dimension for the unparticle instead of two, due to the fact that the square of the mass of the AdS scalar field $\Phi$ is positive\footnote{In AdS, a stable particle is allowed to have negative mass squared as long as it satisfies the so-called Breitenlohner-Freedman bound.}. As a result, the
unparticle propagator has zeros defined by
$G_U(0)=0$, not infinities. This is the principal result of this construction.   

\begin{figure}
\begin{center}
\includegraphics[width=3.2in]{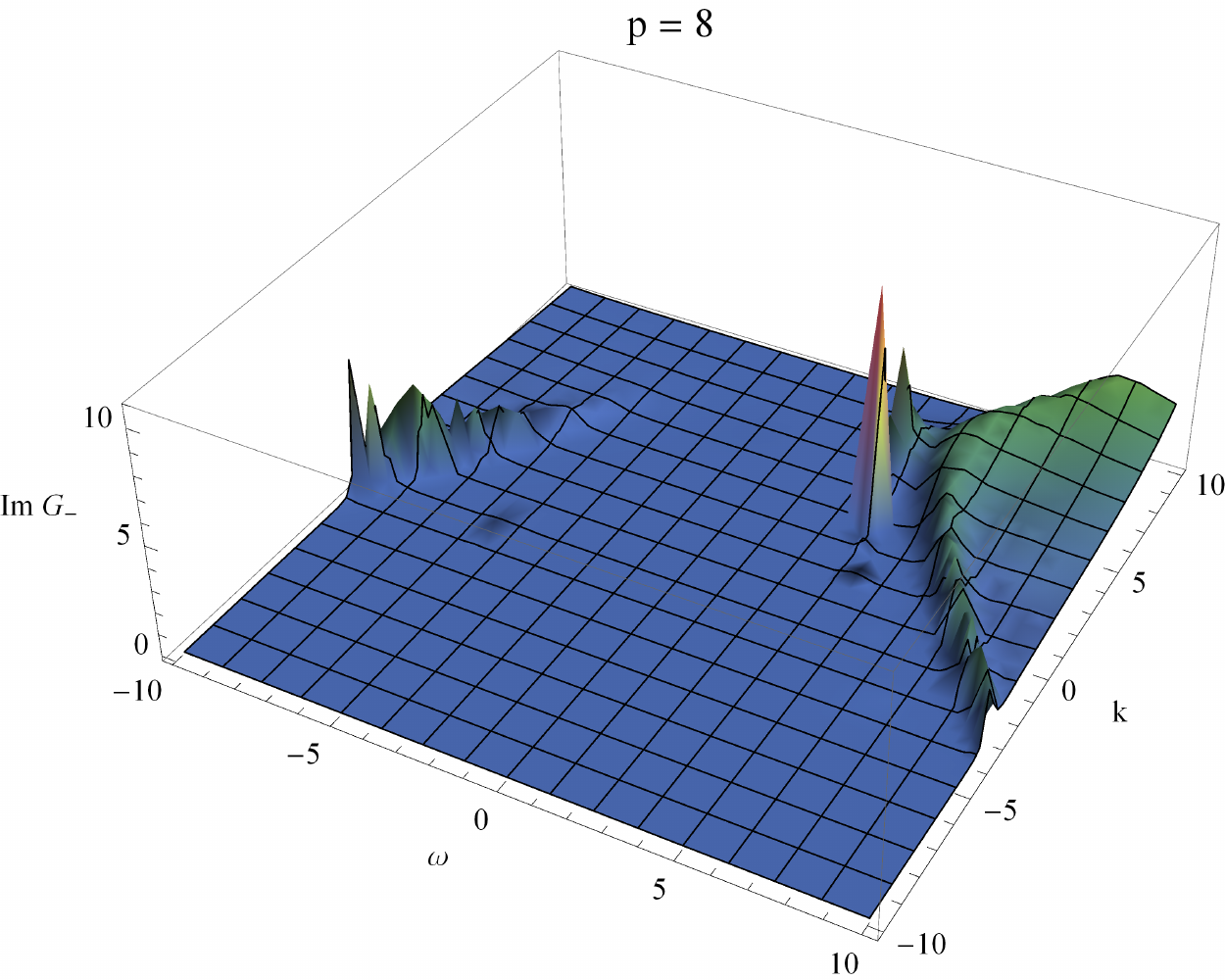}
\caption{  Imaginary part of the spectral function for a Pauli coupling of $p=8$.  The absence of spectral weight for a range of energy flanking $\omega=0$ results from the zeros-poles duality.  }
\label{gap}
\end{center}
\end{figure}
\begin{figure}
\begin{center}
\includegraphics[width=3.0in]{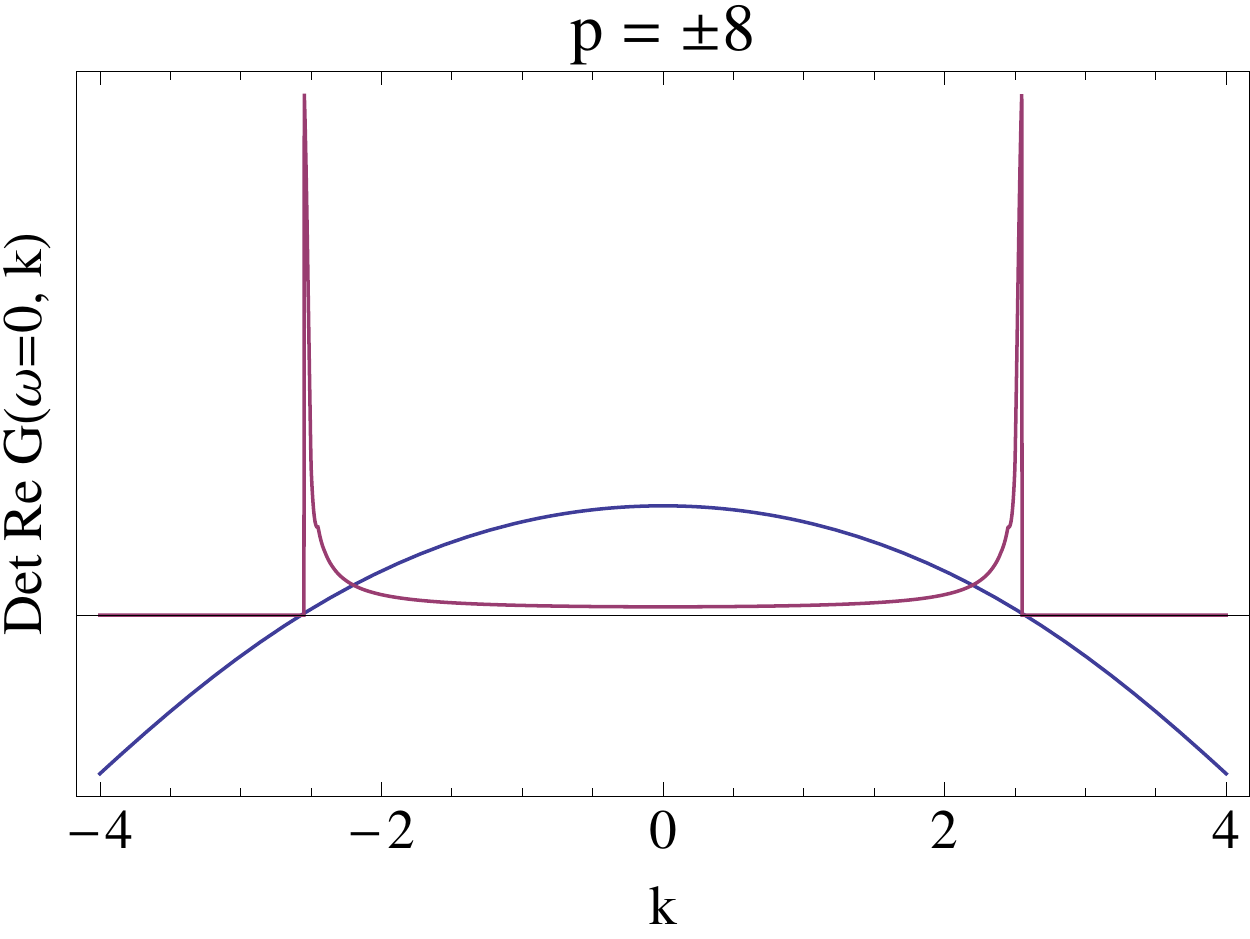}
\caption{  Duality between poles and zeros simply changing the sign of the Pauli term.  For negative values of $p$ the Green function exhibits poles.  By duality, the poles are converted into zeros when $p$ changes sign and becomes positive.  It is the duality that underlies the suppression of the spectral weight and the opening of a gap once $p$ exceeds a critical value.}
\label{zpduality}
\end{center}
\end{figure}

 
The unparticle construction of propagators that have zeros is made possible because the mapping to anti de Sitter space fixed the scaling dimension.  What this suggests is that within the general framework of the gauge-gravity duality, it should be possible to add various terms to the bulk gravity model which couple to the scaling dimension of the dual fermionic operators at the boundary to engineer propagators that exhibit zeros.  Indeed, such a construction is possible\cite{edalati,alsup,garrett}.  Computing fermionic correlators at the boundary requires fermionic fields in the bulk described by an appropriate equation of motion.  All the early work on fermionic correlators\cite{zaanen,faulkner} in AdS was based on probe fermions obeying a Dirac action
\beq\label{sdirac}
S=(\psi,\bar\psi)=\int d^dx\sqrt{-ig}i\bar\psi(\slashed{D}-m+\cdots)\psi
\eeq
in which the mass was varied.  The asymptotic solutions ($r\rightarrow \infty$)  to the equations of motion for $\psi$ scale $a r^{m}$ and $br^{-m}$.  One is free to interpret either term as the source or the response.  In the standard quantization, the response term vanishes at the boundary.  Consequently, the associated boundary fermionic propagator is the ratio $b/a$.  The ``$\cdots$'' in the action for the probe fermions reveals that bulk gravity theory is not unique because the boundary theory is not known, just the correlators.  Such correlators all possess Fermi surfaces, though the excitations can even be of the marginal Fermi liquid type\cite{zaanen,faulkner}.  However, when the boundary theory is specified, thereby fixing the gravity theory, no Fermi surfaces arise\cite{gauntlett,gubser,gubser2}.  All the spectral functions\cite{gauntlett,gubser,gubser2} are of the unparticle kind with a vanishing spectral weight at the chemical potential.  What to make then of the bottom-up constructions based on the Dirac equation? Recall, a restriction for the implementation of the gauge-gravity duality is that the boundary theory has to have a finite coupling constant, strictly absent from a Fermi liquid\cite{polchinski}.  This can be fixed by specifying extra details of the bulk theory; that is, filling in the ``$\cdots$'' in the probe fermion action in Eq. (\ref{sdirac}).   There are many terms that can be added, all of which involve higher derivative operators.  That such terms are not the leading operators in the bulk theory does not imply that they cannot change the boundary dynamics.  A possible mechanism is that they change the scaling dimension of the boundary correlators.  One such operator which does this is the Pauli term.  Consider replacing the ``$\cdots$'' in Eq. (\ref{sdirac}) by $-ip[\Gamma^\mu,\Gamma^\nu]F_{\mu\nu}$ where $\Gamma^\mu$ is a Dirac matrix.  Such a term is well known to give rise to the anomalous magnetic moment of the electron in flat space.  Even if the details of the bulk spacetime vary from Reissner-Nordstrom\cite{edalati,alsup} to Schwarzschild\cite{garrett}, as long as the spacetime at the boundary is asymptotically anti de Sitter, the fermionic propagator is gapped\cite{edalati,garrett,alsup} as shown in Fig. (\ref{gap}) once $p$ exceeds a critical value.  The origin of the gap is now well understood\cite{alsup,garrett} and stems from an exact duality
\beq
{\rm Det \Re G^R}(\omega=0,k;p)=\frac{1}{{\rm Det \Re G^R}(\omega=0,-k;-p)}\nonumber\\,
 \eeq
between zeros and poles as first shown by Alsup,  et al.\cite{alsup} for the Reissner-Nordstrom spacetime and by Vanacore and Phillips\cite{garrett} for the Schwarzschild case.  For sufficiently negative values of $p$, only poles exist.  This means that if we flip the sign of $p$, all the poles must be converted into zeros.  Zeros imply a gap and hence a boundary theory at finite coupling.  Consequently, the Pauli term appears to be a fundamental aspect of the physics of boundary correlators in AdS as partly evidenced by the fact that it is present in top-down constructions\cite{gauntlett, gubser,gubser2}.  An example of the zeros poles duality is depicted in Fig. (\ref{zpduality}).  The mechanism for the conversion of poles to zeros is that the Pauli term couples to the scaling dimension of the probe fermions.  While this is also true for the mass term,  $\rm Det \Re G^R$ is restricted in this case to be $\pm 1$.  No such restriction applies when the Pauli term is present.  

\section{Superconducting Instability}

Because unparticles do not have any particular energy, they should be
useful in describing physics in which no coherent quasiparticles
appear, as in the normal state of the cuprates.  As a result, we have advocated then that unparticles provide a description of the incoherent fermionic 
contribution to the charge density in the pseudo gap and strange metal regions of the cuprates\cite{unparticles}.  Since all
formulations of superconductivity start with well-defined
quasiparticles, we explore what happens when we use a quasiparticle spectral function with a scaling form indicative of unparticles. 
Such an approach is warranted given that the cuprates
 exhibit a color change\cite{marel1,rubhaussen} upon a transition to
 the superconducting state as
 evidenced most strikingly by the violation of the
 Ferrell-Glover-Tinkham sum rule\cite{marel1}. Some initial work along these lines has been proposed
previously\cite{sczhang,balatsky,andersonchak}, which have all been
based on the Luttinger-liquid Green function. As remarked earlier, the
advantage of the unparticle approach is that it is completely general
regardless of the spatial dimension unlike the Luttinger-liquid
one which must be restricted to $d=1+1$.  To obtain the general
result, we work first with the scaling form of the spectral function as in
Eq. (\ref{specfun}), generalizing the procedure in \cite{balatsky}.   Complementary to this work is a recent paper in which the pairing itself, rather than the propagators, was treated as unparticle-like\cite{leblanc}.  

The equation for the existence of an instability
in terms of the Green function is
\begin{eqnarray}
1 & = & i\lambda\sum_{\vec{k}}\left|w_{\vec{k}}\right|^{2}G\left(\vec{k}+\vec{q}\right)G\left(-\vec{k}\right).
\end{eqnarray}
This gives the zero temperature result. We take the interaction strength
$\lambda$ as a constant of mass dimension $2-d$, and $w_{\vec{k}}$ as
a filling factor. We work in the center of mass frame, such that $\vec{q}=\left(q_{0},0\right)$.
The critical temperature for a second-order transition corresponds
to $q_{0}=0$. Then switch to imaginary time to work at finite temperature,
so that the new equation reads
\begin{eqnarray}
\label{crittempconstraint}
1 & = & \lambda T\sum_{n,\vec{k}}\left|w_{\vec{k}}\right|^{2}G\left(\omega_{n},\vec{k}\right)G\left(-\omega_{n},-\vec{k}\right).
\end{eqnarray}
The Green function is related to the spectral function via

\begin{eqnarray}
G\left(\omega_{n},\vec{k}\right) & = & \int_{-\infty}^{\infty}dx\frac{A\left(x,\vec{k}\right)}{x-i\omega_{n}}.
\end{eqnarray}
Then we obtain, using $\omega_{n}=\pi T\left(2n+1\right)$,

\begin{eqnarray}
1 & = & \frac{\lambda}{2}\int dxdy\sum_{\vec{k}}\left|w_{\vec{k}}\right|^{2}A\left(x,\vec{k}\right)A\left(y,-\vec{k}\right)\nonumber\\
&&\times\frac{\tanh\left(x/2T\right)+\tanh\left(y/2T\right)}{x+y}.
\end{eqnarray}
The $\vec{k}$-dependence is recast in terms of $\xi\left(\vec{k}\right)$, which
is in general some function of $\vec{k}$ with units of energy, which for
instance can always be done for an isotropic system. In BCS, $\xi\left(\vec{k}\right)$
would correspond to kinetic energy. Take $\sum_{\vec{k}}\left|w_{\vec{k}}\right|^{2}\to \left(\text{Volume}\right)^{-1}\times N\left(0\right)\int d\xi$
so we pull out a constant density of states. The integral is now rewritten
as
\begin{eqnarray}
1 & = & \frac{g}{2}\int dxdy\int_{0}^{\omega_{c}}d\xi A\left(x,\xi\right)A\left(y,\xi\right)\nonumber\\
&&\times\frac{\tanh\left(x/2T\right)+\tanh\left(y/2T\right)}{x+y}
\end{eqnarray}
where $\lambda N\left(0\right)\times\left(\text{Volume}\right)^{-1}=g$ such that $g$
is a dimensionless measure of interaction strength. Evaluating the integral at any temperature gives the minimal coupling
to cause a pairing instability, and this equation traces out a phase diagram for $g$ and $T$. 

To obtain qualitative information regarding the role that scale invariance plays in the BCS instability, let us  impose the scaling
form at the outset.  Then approximately

\begin{eqnarray}
1 & = & \frac{g}{2}\tilde{T}^{2\left(1+\alpha_A\right)}\int dxdy\int_{0}^{\omega_{c}/\tilde{T}}d\xi A\left(x,\xi\right)A\left(y,\xi\right)\nonumber\\
&&\times\frac{\tanh\left(x/2W\right)+\tanh\left(y/2W\right)}{x+y}
\end{eqnarray}
where the tilde denotes the ratio of that energy to $W$, \emph{e.g.} $\tilde{T}\equiv\frac{T}{W}$.
The scaling form of the spectral function confers a scaling form for $g$ like

\begin{eqnarray}
g\left(\tilde{T},\tilde{\omega}_{c}\right) & = & \tilde{T}^{-2\left(1+\alpha_A\right)}f_g\left(\frac{\tilde{T}}{\tilde{\omega}_{c}}\right).
\end{eqnarray}
Now sequentially we take a logarithm, derivative and finally rescale
the remaining integral back to obtain

\begin{eqnarray}
\frac{dg}{d\ln\tilde{T}} & = & -2\left(1+\alpha_A\right)g+\frac{g^{2}}{2}\omega_{c}\int dxdyA\left(x,\omega_{c}\right)A\left(y,\omega_{c}\right)\nonumber\\
&&\times\frac{\tanh\left(x/2T\right)+\tanh\left(y/2T\right)}{x+y}.
\end{eqnarray}
The second term is positive-definite. This term can conceivably be
small if there is relatively little spectral weight near $\omega_{c}$
within the scaling form or, equivalently, that $g$ is not very susceptible to changes in $\omega_c$. In this event,  we have 
\beq
\frac{dg}{d\ln\tilde{T}} =  -2\left(1+\alpha_A\right)g+O\left(g^{2}\right).
\label{betafcn}
\eeq
The right-hand side of this expression is strictly negative for our region of interest where $\alpha_A>0$.  Hence, we find quite generally that the critical temperature increases as
the coupling constant decreases! 

\begin{figure}[h!]
\begin{center}
\includegraphics[width=3.0in]{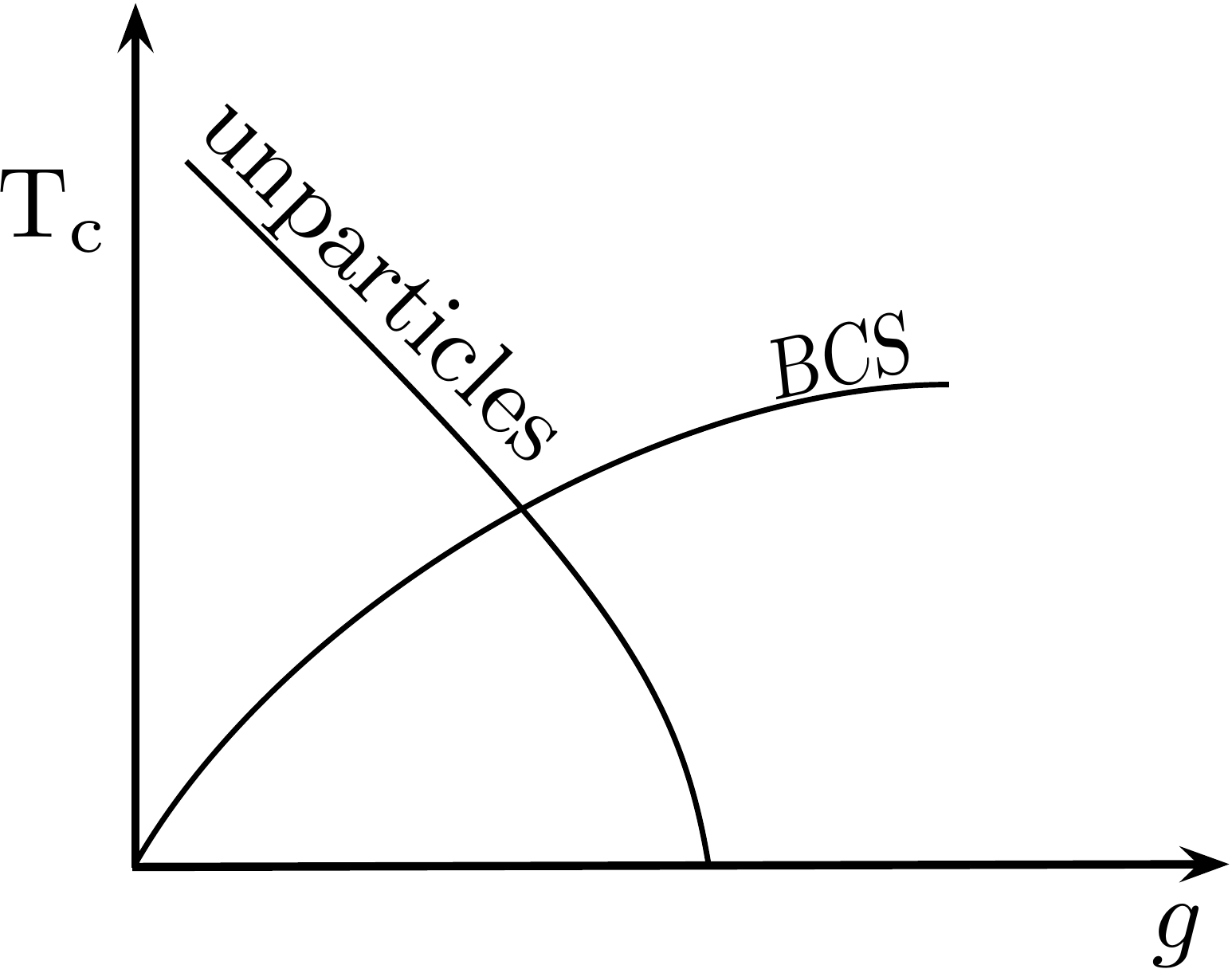}
\caption{  Plot of the $\beta$-function for the superconducting
  transition in the ladder approximation for unfermions,
  Eq. (\ref{betafcn}).  We have considered the case $d_U>d/4$ for the scaling dimension
  of the unparticle field, a condition naturally satisfied in the AdS construction of unparticles.  The contrast with the standard prediction
  of BCS theory is shown.   The dome-shaped superconducting region is highly suggestive that a two-fluid model of unparticles and pole-like excitations underlies
  superconductivity in the cuprates.}
\label{TC}
\end{center}
\end{figure}

This stands in stark contrast to the Fermi
liquid case in which just the opposite state of affairs obtains.  
This is illustrated clearly in Fig. (\ref{TC}).  In the context of the
cuprate superconductor problem, the opposing trends for $T_c$ versus
the pairing interaction suggests that perhaps a two-fluid model
underlies the shape of the superconducting dome assuming, of course, that a similar behaviour for $T_c$ as a function of doping persists.\footnote{There is of course no connection between $g$ and the hole-doping level, $x$. Certainly a generalization of this work to a doped model that admits unparticles could be studied as a function of doping to see if the qualitative trends in Fig. (\ref{TC}) obtain.} Since the
transition to the superconducting state breaks scale invariance, the
particle picture should be reinstated.  Consequently, we expect the
broad spectral features dictated by the branch cut of the unfermion
propagator to vanish and sharp quasiparticle features to appear upon
the transition to the superconducting state as is seen experimentally\cite{he}.  

\section{Closing}

What I have laid out here is a way of thinking about physics when the particle concept breaks down.
The Luttinger count ensures that a purely IR limit, or particle interpretation, of the 
number of charges exists even if interactions are present.  However, as we have  shown, when zeros are present, no such theorem
applies.  Zeros obtain from a divergent self energy and Eq. (\ref{nlutt}) is not valid in this limit.  In fact, this was pointed out explicitly by Dzyaloshinskii\cite{dzy} but that such a state of affairs obtains in a Mott insulator was overlooked.  Zeros indicate then that some charged stuff exists that has no particle interpretation.   I have illustrated here how unparticles can be used to describe such a breakdown and dynamical generation of a mass gap in bottom-up AdS constructions is consistent with the zeros picture.  Ultimately, a scale invariant sector must arise from a non-trivial fixed point.  Establishing that such a non-trivial fixed point exists at strong coupling remains the outstanding problem in this field.  

\textbf{Acknowledgements}These lectures were based on a series of papers with B. Langley, K. Dave, J. Huttasoit, C. Kane, R. Leigh, and M. Edalati.
This work is supported by NSF DMR-1104909 which grew out of earlier work funded by the
Center for Emergent Superconductivity, a DOE Energy Frontier Research Center, Grant No.~DE-AC0298CH1088. 

\appendix{Appendix}

We show here that if only if a linear relationship between the unparticle and particle fields is maintained (as in Eq. (\ref{uphiphi})), then a Gaussian action for the unparticles obtains.
Let us turn the action in terms of the massive fields into an action
in terms of unparticle fields. The original partition function is
given by
\begin{eqnarray*}
\mathcal{Z} & = & \int\mathcal{D}\phi_{n}e^{i\int d^{d}p\mathcal{L}\left[\left\{ \phi_{n}\right\} \right]}\\
\mathcal{L} & = & \frac{1}{2}\sum_{n}B_{n}\phi_{n}\left(p\right)\left(p^{2}-M_{n}^{2}\right)\phi_{n}\left(-p\right).
\end{eqnarray*}
where $n$ is to indicate a sum over the mass $M_{n}$. This sum can remain a general sum over various free fields, but we will ultimately take the limit where the sum is a continuous sum over all masses. The factor $B_n$ is a weight factor that, in the continuous mass limit, will change the mass dimension of $\phi_n$. We introduce
a Lagrange multiplier through a factor of unity and simply integrate
over the all fields that are not $\phi_{U}$ to obtain
\begin{eqnarray*}
\mathcal{Z} & = & \int\mathcal{D}\phi_{n}\mathcal{D}\phi_{U}\mathcal{D}\lambda\exp\left\{ i\int d^{d}p\left(\frac{1}{2}\sum_{n}B_{n}\left(p^{2}-M_{n}^{2}\right)\phi_{n}^{2}+\lambda\left(\phi_{U}-\sum_{n}F_{n}\phi_{n}\right)\right)\right\} \\
 & = & \int\mathcal{D}\phi_{U}\mathcal{D}\lambda\exp\left\{ i\int d^{d}p\left(\lambda\phi_{U}-\frac{1}{2}\lambda^{2}\sum_{n}\frac{F_{n}^{2}}{B_{n}\left(p^{2}-M_{n}^{2}\right)}\right)\right\} \\
 & = & \int\mathcal{D}\phi_{U}\exp\left\{ \frac{i}{2}\int d^{d}p\phi_{U}\left(p\right)\left(\sum_{n}\frac{F_{n}^{2}}{B_{n}\left(p^{2}-M_{n}^{2}\right)}\right)^{-1}\phi_{U}\left(-p\right)\right\} 
\end{eqnarray*}
with repeated absorptions of normalization contants into the measure. The factor $F_n$ is another weight factor, this time chosen to determine the scaling dimension of the unparticle field $\phi_U$. Bescause $F_n$ is chosen to give $\phi_U(x)$ a scaling dimension $d_U$, in the continuous mass limit the ratio $F_n^2/B_n \sim \left(M_n^2\right)^{d_U-\frac{d}{2}}$. This is necessary because of how $F_n$ imposes the scaling dimension.
Hence we identify the propagator of the unparticle field as
\begin{eqnarray*}
G_{U}\left(p\right) & = & \sum_{n}\frac{F_{n}^{2}}{B_{n}\left(p^{2}-M_{n}^{2}\right)}\\
 & \sim & \left(p^{2}\right)^{d_{U}-\frac{d}{2}}.
\end{eqnarray*}

This argument can also be run in reverse.  Namely, if we assume a Gaussian action for the unparticles then the Lagrange multiplier constraint in the form of Eq. (\ref{uphiphi}) is implied.

\bibliographystyle{ws-procs9x6}

\end{document}